\documentclass[twocolumn,showpacs,preprintnumbers,amsmath,amssymb]{revtex4}

\usepackage{float}
\usepackage{graphicx}
\usepackage{dcolumn}
\usepackage{bm}
\usepackage[dvips]{color}
\definecolor{grau}{rgb}{0.9,0.9,0.9}
\definecolor{hellgrau}{rgb}{0.935,0.935,0.935}
\definecolor{gelb}{rgb}{1,1,0.7}
\definecolor{rot}{rgb}{1,0.875,0.935}
\definecolor{blau}{rgb}{0.1,0.21,0.5}
\definecolor{gruen}{rgb}{0.25,0.5,0.35}
\definecolor{hellgelb}{rgb}{0.4,0.5,0.5}
\definecolor{dgelb}{rgb}{0.6,0.5,0.5}
\definecolor{dunkelgruen}{rgb}{0.0,0.45,0.0}
\definecolor{dunkelrot}{rgb}{0.5,0.06,0.05}
\definecolor{dt}{rgb}{0.75,0.65,1}
\definecolor{tuerkis}{rgb}{0.95,0.9,1}

\begin{document}

\preprint{APS/123-QED}

\title{Electric Characteristics of Rotational States positive parity in isotopes $^{170,172,174}Yb$ }

\author{P.N.Usmanov}%
 \email{usmanov1956.56@mail.ru}
\affiliation{%
Institute of Nuclear Physics, Academy of Science of Uzbekistan,
100214, Ulugbek, Tashkent, Uzbekistan
}%

\author{A.A.Okhunov }
 \email{Corresponding author: aaokhunov@gmail.com}\homepage{http://www.iium.edu.my}
\affiliation{ Department of Science in Engineering, KOE,
International Islamic University Malaya, P.O Box 10, 50728
Kuala Lumpur, Malaysia }%

\affiliation{%
Institute of Nuclear Physics, Academy of Science of Uzbekistan,
100214, Ulugbek, Tashkent, Uzbekistan }%



\author{H. Abu Kassim}
 \email{hasank@um.edu.my}
\affiliation{ Department of Physics, University of Malaya, 50603
Kuala Lumpur, Malaysia }%

\author{U.S. Salikhbaev}
 \email{salikhbaev@inp.uz}
\affiliation{ Institute of Nuclear Physics, Academy of Science of
Uzbekistan, 100214, Ulugbek, Tashkent, Uzbekistan }%

\date{\today}

\begin{abstract}
Accounting for Coriolis mixing of experimentally known rotational
bands with $K^{\pi}< 3^+$, non-adiabatic effects in energy and
electric characteristics of excited states are investigated,
within phenomenological model.

The energy and wave function structure of excited states are
calculated. The finding reveals that the bands mixing has been
found to have considerable impact on the wave function of
low-lying states $0^+$ and $2^+$ bands.

In addition, the probabilities of $E2$-- transitions
have been calculated. The values from calculations of $B(E2)$--
transitions from $0_2^+$, $0_3^+$, $2_1^+$, and $2_2^+$ bands
are compared with the experimental data.

\end{abstract}



\maketitle

\section{\label{sec:level1}Introduction }

Despite the fact that the structure of deformed nuclei and nature
of low excited levels have been substantially studies over more
than four decades, this still occupies a central part of today's
research \cite{Bohr}-\cite{Burke}.

An extensive research interest in the properties of deformed
nuclei has risen in recent years with the exploration of a new
collective isovector magnetic dipole mode \cite{Bohle, Zilges}.
The measured values of excited energy of magnetic mode are found
to be not so high in an excited spectrum, and consideration of
mixing with low-lying exciting states appear to lead to an
interesting physical phenomena \cite{Usmanov, Usmanov2010}.

The nuclei $^{170,172,174}Yb$ have been well studied. It is
important to note that these are investigated in a number of ways
such as radioactive decay of $^{170,172,174}Lu$, and different
nuclear reactions. In these isotopes, many $1^+$ states and
$K^{\pi}=0^{+}, 2^{+}$ bands have been observed. For instance, the
excited energy $K^{\pi}=0^{+}$ and $2^{+}$ it rises with the
increase in number of neutrons (see Figure 1).

\begin{figure}[h!] \label{Fig.1}
  \includegraphics[width=0.45\textwidth]{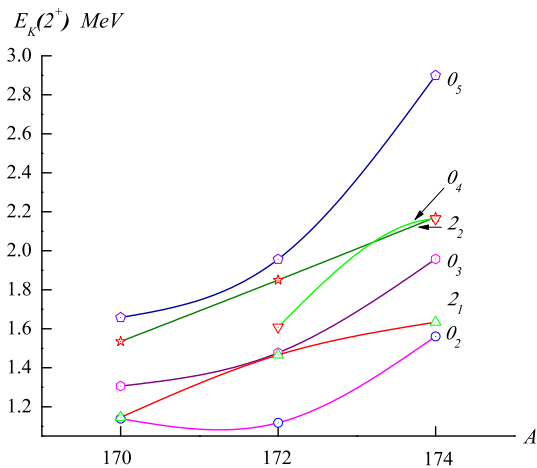}\\
  \caption{(Color online) The energy of $I=2$ states of
$K^{\pi}=0_2^+, 0_3^+, 0_4^+, 0_5^+, 2_1^+$ and $2_2^+$ bands
  in isotopes $^{170,172,174}Yb$. }
\end{figure}

The values of probability of $B(E2)$ with low-lying levels of
$K^{\pi}=0^{+}$, $2^{+}$ bands, and also Rasmusson's parameter
value $X_I(E0/E2)$, dimensionless units matrix element $\rho(E0)$
for the $E0$-- transition and a multipole mixture coefficients
$\delta(E2/M1)$ are defined experimentally
\cite{Baglin}-\cite{Browne}.

Numerous conducted experiments on defining spectroscopic
characteristics of low-lying exciting states, particular
$K^{\pi}=1^{+}$ in deformed nuclei \cite{Zilges}, have motivated
the further theoretical investigations. In this case,
investigations influence of $K^{\pi}=1^{+}$ states to the
properties of low-lying levels is actual.

Present paper focuses on low-lying states of positive parity of
isotopes $^{170,172,174}Yb$. The calculation are conducted by
utilizing a phenomenological model \cite{Usmanov} which accounts
Coriolis mixture all of the experimentally known low-lying
rotational bands states with $K^{\pi}< 3^{+}$.

Experimentally observed $K$-- forbidden transitions as well as
nonadiabaticities of energy and in ratios of $E2$-- transitions
can be explained by Coriolis mixture states.

\section{\label{sec:level2}The Model}

To analyze the properties of low-lying positive parity states in
$Yb$ isotopes, the phenomenological model of \cite{Usmanov} is
exploited. This model takes into account the mixing of states of
the $K^{\pi}=0^+, 2^+$ and $1^+$ bands. The Hamiltonian model is
\begin{eqnarray}\label{Eq.1}
H=H_{rot}(I^2)+ H_{K,K'}
\end{eqnarray}

\noindent
\begin{eqnarray}\label{Eq.2}
H_{_{K'K}}^{\sigma}(I)=\omega_{_K}\delta_{_{K,K'}}-\omega_{rot}(I)(j_x)_{_{K,K'}}\zeta(I,K)\delta_{_{K,K'\pm1}}
\end{eqnarray}

\noindent where $\omega_{_K}$-- bandhead energy of rotational
band, $\omega_{rot}(I)$-- an angular frequency of rotational
nucleus, $(j_x)_{_{K,K'}}$-- matrix elements which describe
Coriolis mixture between rotational bands and
\begin{eqnarray}
\zeta(I,0)=1 \ \ \ \ \
\zeta(I,2)=\left[1-\frac{2}{I(I+1)}\right]^{\frac{1}{2}} \nonumber
\end{eqnarray}

The eigenfunction of Hamiltonian model (\ref{Eq.1}) is
\begin{eqnarray}\label{Eq.3}
|IMK>&=&\sqrt{\frac{2I+1}{16\pi^2}}\left\{\sqrt{2}\Psi_{gr,K}^ID_{MK}^{I}(\theta)\right.
\\ \nonumber
&+&\left.\sum_{K'}\frac{\Psi_{K',K}^I}{\sqrt{1+\delta_{K',0}}}
\left[D_{M,K'}^I(\theta)b_{K'}^+\right.\right.
\\ \nonumber
&+&\left.\left.(-1)^{I+K'}D_{M,-K'}^I(\theta)b_{-K'}^+\right]\right\}|0>
\nonumber
\end{eqnarray}

\noindent here $\Psi_{K',K}^{I}$ is the amplitude of mixture of
basis states.

The rotational part of Hamiltonian (\ref{Eq.1}) $H_{rot}(I)$ is
diagonal by wave functions (\ref{Eq.3}). Note that $H_{rot}(I)$ is
determined by exploiting Harris parameterization for energy and
angular momentum \cite{Harris}
\begin{eqnarray}\label{Eq.4}
E_{rot}(I)=\frac{1}{2}\Im_0\omega_{rot}^2(I)+\frac{3}{4}\Im_1\omega_{rot}^4(I)
\end{eqnarray}
\begin{eqnarray}\label{Eq.5}
\left[I(I+1)\right]^{1/2}=\Im_0\omega_{rot}(I)+\Im_1\omega_{rot}^3(I)
\end{eqnarray}

\noindent where $\Im_0$ and $\Im_1$-- are the inertia parameters
of the rotational core.

The rotational frequency of the core $\omega_{rot}(I)$ is found by
solving cubic equation (\ref{Eq.5}). This equation has two
imaginary roots and one real root. The real root is as follows
\begin{eqnarray}\label{Eq.6}
\omega_{rot}(I)&=&
\left\{\frac{\widetilde{I}}{2\Im_1}+\left(\left(\frac{\widetilde{I}}{2\Im_1}\right)^2+
\left(\frac{\Im_0}{3\Im_1}\right)^3\right)^{\frac{1}{2}}\right\}^{\frac{1}{3}}
\\ \nonumber
&+&\left\{\frac{\widetilde{I}}{2\Im_1}-\left(\left(\frac{\widetilde{I}}{2\Im_1}\right)^2+
\left(\frac{\Im_0}{3\Im_1}\right)^3\right)^{\frac{1}{2}}\right\}^{\frac{1}{3}}
\end{eqnarray}

\noindent where $\widetilde{I}=\sqrt{I(I+1)}$. Equation
(\ref{Eq.6}) gives $\omega_{rot}(I)$ at the given spin $I$ of the
core.

 Solving the Shr\"{o}dinger equation
\begin{eqnarray}\label{Eq.7}
H_{K,K'}^{\sigma}\Psi_{K,K'}^I=\mathcal{E}_K(I)\Psi_{K,K'}^I.
\end{eqnarray}

\noindent we define eigne function and energy of a Hamiltonian.
The total energy of state is defined by
\begin{eqnarray}\label{Eq.8}
E_{K}(I)=E_{rot}(I)+\mathcal{E}_K(I)
\end{eqnarray}

\subsection{\label{sec:level2}Energy spectra and structures of the states}

The calculations have been carried out for the isotopes
$^{170,172,174}Yb$. All experimentally known rotational bands of
positive parity with $K^{\pi}< 3^+$ have been included in basis
Hamiltonian states.

The experiment suggests that $m=5$ band with $K^{\pi}=0_m^+$, one
band $\ell=1$ with $K^{\pi}=2_{\ell}^+$, and $\nu=19$ with
$K^{\pi}=1_{\nu}^+$ states in $^{170}Yb$ \cite{Baglin}. These all
$n=m+\ell+\nu=25$ rotational bands have been included in the basis
states of Hamiltonian (\ref{Eq.1}). For the isotopes
$^{172,174}Yb$, basis states of Hamiltonian include $n=15$ ($m=5$,
$\ell=2$ and $\nu=8$) and $n=22$ ($m=5$, $\ell=2$ and $\nu=15$),
correspondingly \cite{Zilges, Singh, Browne}.

The parameters of inertia $\Im_0$ and $\Im_1$ are estimated by
exploiting Harris parameterization (\ref{Eq.4}), and using the
experimental data for energy up to spin $I\leq 8 \hbar$ for ground
band \cite{Okhunov}.

The Hamiltonian (\ref{Eq.2}) has transformational properties, that
the state (\ref{Eq.3}) can be classified as quantum number ---
$\sigma=\pm 1$ signature, which imposes restrictions on angular
momentum values.
$$
(-1)^I\sigma=1.
$$

For the states with negative signature $\sigma=-1$, Hamiltonian
(\ref{Eq.2}) has dimension $n=\ell+\nu$, as in bands with
$K^{\pi}=0_{m}^{+}$ the are no condition states with odd spins
$I$. For the states with positive signature $\sigma=+1$,
Hamiltonian (\ref{Eq.2}) has dimension $n=m+\ell+\nu$.

The model parameters are described as follows:\\

a) the bandhead energy ground $\omega_{0_1}$ and $K^{\pi}=0_m^+$
bands has taken from experiment, as they are not revolted by
Coriolis force. Bandhead energy of $1_{\nu}^+$ bands are also
defined from an experiment \cite{Zilges, Baglin}
$$\omega_{1_{\nu}}=E_{1_{\nu}}^{exp.}(I=1)-E_{rot}(I=1);$$

\noindent b) matrix elements
$(j_x)_{2_{\ell}1}=(j_x)_{2_{\ell}1_{\nu}}$ and bandhead energy of
$2_{\ell}^+$-- bands $\omega_{2_{\ell}}$ are determined from the
most favored experimental and theoretical spectrum of energy
states with a negative signature $\sigma=-1$, e.a. for energy
state for even spins $I$;

\noindent c) the matrix elements
$(j_x)_{0_{m}1}=(j_x)_{0_{m}1_{\nu}}$ defined by the least square
method from the best fitted of theoretical energy spectra state
with positive signature $\sigma=+1$ with experimental data.

The obtained values of model parameters are presented in Table 1.

\begin{widetext}
\begin{center}
\begin{table} [h]{\textbf{\caption {Parameters used in the calculations for the isotopes $^{170,172,174}Yb$}}}
\begin{small}
\begin{tabular*} {0.79\textwidth}%
{@{\extracolsep{\fill}}cccccccc} \hline\hline \ \ $A$ \ \ & \ \

   \ \ $(j_x)_{0_1, 1_{\nu}}$ \ \ &  \ \ $(j_x)_{0_2, 1}$ \ \ &
   \ \ $(j_x)_{0_3, 1}$ \ \ &  \ \ $(j_x)_{0_4, 1}$ \ \ &  \ \ $(j_x)_{0_5, 1}$ \ \ &
   \ \ $(j_x)_{2_1, 1}$ \ \  &    \ \ $(j_x)_{2_2, 1}$ \ \   \\ \hline
 $170$ &  0.1864 & 0.3936 & 0.6586 & 0.9081 & 0.0009 &
0.7278 & --  \ \\
 $172$ &  0.2754 & 0.9777 & 0.7176 & 0.11 & 0.30 & 0.325 &
0.21 \ \\
 $174$ &  0.185 & 0.4 & 0.25 & 0.15 & 0.20 & 0.085 &
0.1 \  \\ \hline\hline
\end{tabular*}
\end{small}

\vspace{0.5cm} \footnotesize{Note: $(j_x)_{_{K', K}}$-- are matrix
elements of the Coriolis interactions.}
\end{table}
\end{center}
\end{widetext}

Calculation comparison of energy with experimental values for
$^{170,172,174}Yb$ is illustrated in Figures 2,3 and 4,
correspondingly.

Apparently, one may see from the Figures that the model
qualitatively reproduces experimental energy of rotational states
up to energy $3 MeV$. However, in high spin values $I\geq 12\hbar$
noticeable deviation has been observed in calculated values of
energy and that obtained from experiment. Note that this deviation
increases with the growth of angular momentum $I$. This is
probably due to the fact that the influence of rotation on
internal nuclei structure has not been considered in this model.
In future, we will study electromagnetic properties of low-lying
states $I<10\hbar$.

Amplitude of the states $\Psi_{K^{'},K}$ for $K^{\pi}=2_{1}^{+}$--
and $0_{2}^{+}$-- bands for $^{170,172,174}Yb$, are provided in
Figures 5, 6, 7, 8, 9 and 10, respectively. The components which
have small values are not illustrated in Figure. Also the
components $\Psi_{1_{\nu},K}$ band are not given except for the
first $1_1^+$. The values for others $\Psi_{1_{\nu}^{+},K}$ states
are define as follows
\begin{eqnarray}\label{Eq.9}
\Psi_{1_{\nu},K}=\Psi_{1_{1},K}\frac{\omega_{1_1}-
\omega_{K}}{\omega_{1_{\nu}}-\omega_{K}}.
\end{eqnarray}

From Figures 5 and 7, we can see that $K^{\pi}=0_2^+$ and
$K^{\pi}=2_1^+$ bands states in $^{170}Yb$, $K^{\pi}=0_3^+$ and
$K^{\pi}=2_1^+$ bands in $^{172}Yb$ are mixed strongly even in low
spin values $I$. It is associated with the close location to each
other (see Figure 1). In isotopes $^{170,172}Yb$, considerable
deviation in signature of the states $K^{\pi}=2_1^+$ band can be
observed. This reflects in the values of probability
electromagnetic transitions. In this case, description of quantum
number $K$ is difficult for these states. Thus, in $^{170}Yb$, a
number of research works \cite{Begzhanov} in this context note
that the states with $I=2^+$ (1.1386 MeV) and $I=2^+$ (1.1454 MeV)
$K=2$ and $K=0$, respectively. On the other hand, some works
\cite{Baglin} document that $K=0$ and $K=2$, correspondingly. In
case $^{174}Yb$, the mixture effect is not so strong.

\section{Electric Quadrupole transitions}

With the wave functions calculated by solving the Shr\"{o}dinger
equation (\ref{Eq.7}), reduced probabilities of $E2$-- transitions
between states $I_iK_i$ and states of ground band $I_f0_1$ are
calculated \cite{Usmanov}.
\begin{widetext}
\begin{eqnarray}\label{Eq.10}
B(E2;I_iK_i\rightarrow
I_f0_1)&=&\left\{\sqrt{\frac{5}{16\pi}}eQ_0\left[\Psi_{0_1,0_1}^{I_f}
\Psi_{0_1,K_i}^{I_i}C_{I_i0;20}^{I_f0}+\sum_{n}\Psi_{K_n,0_1}^{I_f}
\Psi_{K_n,K_i}^{I_i}C_{I_iK_n;20}^{I_fK_n}\right]\right. \nonumber \\
&+&\left.
\sqrt{2}\left[\Psi_{0_1,0_1}^{I_f}\sum_{n}\frac{(-1)^{K_n}m_{_{K_n}}\Psi_{K_n,K_i}^{I_i}}{\sqrt{1+\delta_{K_n,0}}}
C_{I_iK_n;2-K_n}^{I_f0}+\Psi_{0_1,K_i}^{I_i}\sum_{n}\frac{m_{_{K_n}}\Psi_{K_n,0_1}^{I_f}}{\sqrt{1+\delta_{K_n,0}}}
C_{I_i0;2K_n}^{I_fK_n}\right]\right\}^2
\end{eqnarray}
\end{widetext}

\noindent here $m_{_{K_n}}=<0_1^+|\hat{m}(E2)|K_n^+>$-- is matrix
elements between intrinsic wave functions of ground $(0_1^+)$ and
$K_n^+=0_{m}^+, 2_{\ell}^+, 1_{\nu}^+$ bands which has a value
obtained from experimental data, $Q_0$-- is nuclear intrinsic
quadrupole moment; and $C_{I_iK_i;2(K_i+K_1)}^{I_fK_f}$--
Clebsch-Gordan coefficient.

For the reduced probabilities of $E2$-- transitions from the $I=2$
state we have following equation in adiabatic approximation:
\begin{eqnarray}\label{Eq.11}
&&B^{adia}(E2;2K_n\rightarrow 00_1)= \\ \nonumber
&&=\left(2-\delta_{K_n,0}\right)|m_{_{K_n}}C_{2K_n;2-K_n}^{00}|^2
\end{eqnarray}

\noindent which allows us to identify $m_{_{K_n}}$ parameter from
the experimental data.

The $K^{\pi}=0_2^+$, $0_3^+$ and $2_1^+$ bands are very close to
each other in isotopes $^{170,172}Yb$, which leads to a strong
mixing of states even $I=2$. In this case, the adiabatic
approximation (11) becomes inapplicable to determine $m_{_{K_i}}$.

The magnitude and sign parameters $m_{1_{1}}=m_{1_{\nu}}$ and
$m_{2_{\ell}^+}$ are obtained form the best fitted ratio
probabilities $R_{IK}=B(E2;IK\rightarrow
I+10_1)/B(E2;IK\rightarrow I-10_1)$ and odd states of
$K^{\pi}=2_{\ell}^+$ and $1_{\nu}^+$ bands. In addition, the most
favored ratio $R_{IK}$ and even states (for the positive signature
$\sigma=+1$) help to identify parameters $m_{0_{m}}$.

Table 2 reports $m_{_K}$ parameters which have been used in
calculation $E2$-- transitions.

\begin{widetext}
\begin{center}
\begin{table} [h]
{\textbf{\caption {The values of the parameters $m_K$ and the
intrinsic quadrupole moment $Q_0$, which are used in calculations
(in $efm^2$) }}}
\begin{small}
\begin{tabular*} {0.73\textwidth}%
{@{\extracolsep{\fill}}ccccccccc} \hline\hline
 \ \ $A$ \ \ & \ \
$m_{0_2}$ \ \ &  \ \ $m_{0_3}$ \ \ &
   \ \ $m_{0_4}$ \ \ &  \ \ $m_{0_5}$ \ \ &
   \ \ $m_{1_{\nu}}$ \ \ &  \ \ $m_{2_1}$ \ \ &  \ \ $m_{2_2}$ \ \
   &  \ \ $Q_{0}$\cite{Begzhanov}   \\ \hline
 $170$ & 2  & 24 & 3 & 8 & -5 & 19 & 8 & 780(4)  \\
 $172$ & 10  & 1 & -6.9 & -8 & -5 & 15 & -8  & 791(4)  \\
 $174$ & 8 & 1 & -6.9 & 8 & -10 ($m_{1_1}$=-1.7) & 15 & 8 & 782(4)  \\ \hline\hline
\end{tabular*}
\end{small}
\end{table}
\end{center}
\end{widetext}

Table 3 compares reduced probability $E2$-- transitions with existing experimental data
\cite{Baglin}-\cite{Browne}, \cite{Reich}.  Moreover, reduced matrix elements of $E2$-- transitions for $^{172}Yb$ are provided
in Table 4. In a similar vein, these values are also compared with
experimental values as well as values found by using other models
\cite{Fahlander}-\cite{Arima}.

\begin{widetext}
\begin{center}
\begin{table} [h]
{\textbf{\caption {Reduced probability of $E2$-- transitions in
the isotopes $Yb$ ($e^2fm^4$)}\label{TypoMSP}}}
\begin{small}
\begin{tabular*} {0.90\textwidth}%
{@{\extracolsep{\fill}}ccrr|crr} \hline\hline
  A &  $I_iK_i\rightarrow I_fK_f$  & Exp.  &  Theory  \ &
    $I_iK_i\rightarrow I_fK_f$  & Exp.  &   Theory  \ \\ \hline
$^{170}Yb$ & $22_1\rightarrow 00_1$ & 151(35)\cite{Baglin} & 90 &
 $00_3\rightarrow 20_1$ & 60(15)\cite{Baglin} &  43  \\
& \hspace*{0.5cm}$\rightarrow \ 20_1$ & 269(60)\cite{Baglin} & 60
&  $00_4\rightarrow 20_1$ & 567(118)\cite{Baglin} &  567   \\
& \hspace*{0.5cm}$\rightarrow \ 40_1$ & 27(6)\cite{Baglin} &  10 &
&  &    \\ \hline $^{172}Yb$  & $22_1\rightarrow 00_1$ &
75.6(63)\cite{Singh}; 74.6(57)\cite{Reich}& 82 &
$00_2\rightarrow 20_1$ & 205(60)\cite{Singh} &  100 \\

& \hspace*{0.5cm}$\rightarrow \ 20_1$ & 121(12)\cite{Reich} & 130
& $20_2\rightarrow 00_1$ & 14(1)\cite{Singh}; \ \ \ 14(1)\cite{Reich} &  13 \\

& \hspace*{0.5cm}$\rightarrow \ 40_1$ & 7.3(6)\cite{Singh}; \ \ \
6.8(7)\cite{Reich} & 8.6 &
\hspace*{0.5cm}$\rightarrow \ 20_1$ & 45(7)\cite{Singh}; \ \ \ 52(8)\cite{Reich} &  23  \\

& $42_1\rightarrow 20_1$ & 398(284)\cite{Singh} &  15 &
\hspace*{0.5cm}$\rightarrow \ 40_1$ & 142(20)\cite{Singh}; 140(20)\cite{Reich} & 74  \\

 & \hspace*{0.5cm}$\rightarrow \ 40_1$ & 739(512)\cite{Singh} & 81 &
 $00_3\rightarrow 20_1$ & 0.14(3)\cite{Singh} &  1  \\

& $32_1\rightarrow 20_1$ & 152(11)\cite{Reich} & 154 &
$20_3\rightarrow 00_1$ & 0.4(1)\cite{Singh}; \ \  3.4(2)\cite{Reich} &  3.6  \\

& \hspace*{0.5cm}$\rightarrow \ 40_1$ & 79(6)\cite{Reich} &
73 &  \hspace*{0.5cm}$\rightarrow \ 20_1$ & 0.6(4)\cite{Singh}; \ 11.9(8)\cite{Reich} &  3.0  \\

& $22_2\rightarrow 00_1$ & 20(4)\cite{Singh}; \ \ \
32(4)\cite{Reich} &
 23 & \hspace*{0.5cm}$\rightarrow \ 40_1$ & 1.0(1)\cite{Reich} &  1.2  \\

& \hspace*{0.5cm}$\rightarrow \ 20_1$ & 31(2)\cite{Singh}; \ \ \
51(7)\cite{Reich} &  38 &
$00_4\rightarrow 20_1$ & $>$0.25\cite{Singh} & 48  \\

& \hspace*{0.5cm}$\rightarrow \ 40_1$ & 3.3(4)\cite{Reich} &  2.2
& $20_4\rightarrow 00_1$ & 10(6)\cite{Singh} & 12   \\

& $32_2\rightarrow 20_1$ & 54(7)\cite{Reich} &  42 &
 $00_5\rightarrow 20_1$ &  $>$ 0.27\cite{Singh} & 64  \\

& \hspace*{0.5cm}$\rightarrow \ 40_1$ & 22(3)\cite{Reich} &  21 &
$20_5\rightarrow 00_1$ & 19(9)\cite{Singh} & 21  \\
&  &   & & \hspace*{0.5cm}$\rightarrow \ 20_1$ & $>$ 0.18\cite{Singh} & 23 \\
  \hline
$^{174}Yb$ & $00_2\rightarrow 20_1$ &
$81_{-29}^{+64}$\cite{Browne} & 64 &
 $22_1\rightarrow 20_1$ & 144(30)\cite{Browne} &  133  \\ \hline\hline
\end{tabular*}
\end{small}
\end{table}
\end{center}
\end{widetext}

It is important to note that our results are obtained
consecutively. In the initial step, energy and wave function the
states are computed. Further, by utilizing these wave functions,
reduced probability of $E2$-- transitions are calculated. From the
Table 4, one may gather that performed calculations within our
model provide a better correspondence with experiment data.

\begin{widetext}
\begin{center}
\begin{table} [h]
{\textbf{\caption {Reduced Matrix Elements of $E2$-- transitions
in $^{172}Yb$, calculated within our model which comparison with
experimental data \cite{Fahlander} and are calculated using the
rotational-vibrational model (RVM2) \cite{Faessler} and the IBA-1
 model \cite{Arima} $(eb)$}\label{TypoMSP}}}
\begin{footnotesize}
\begin{tabular*} {1.00\textwidth}%
{@{\extracolsep{\fill}}rrccc|rrccc} \hline\hline
 $I_iK_i\rightarrow I_fK_f$ & Exp. \ &
 \ RVM2 \ & \ IBA-1 \ & Theory &
 $I_iK_i\rightarrow I_fK_f$ & Exp. \ &
 \ RVM2 \ & \ IBA-1 \ & Theory \\ \hline
$20_1\rightarrow 20_1$ & -2.63$_{-0.27}^{+0.28}$ &
 -2.92 & -2.92 & -2.93 &
$20_1\rightarrow 00_1$ & 2.45$_{-0.12}^{+0.12}$ &
2.45$^{a)}$ & 2.45$^{a)}$  & 2.45$^{a)}$ \\
$40_1\rightarrow 40_1$ & -3.54$_{-0.18}^{+0.84}$ & -3.73 & -3.69 &
-3.74 & $40_1\rightarrow 20_1$ & 3.76$_{-0.19}^{+0.19}$ &
3.93 & 3.91 & 3.93 \\
$60_1\rightarrow 60_1$ & -4.31$_{-0.62}^{+0.22}$ & -4.43 & -4.33 &
-4.46 & $60_1\rightarrow 40_1$ & 5.34$_{-0.27}^{+0.27}$ &
4.97 & 4.90  & 4.96 \\
$80_1\rightarrow 80_1$ & -4.49$_{-0.77}^{+0.23}$ & -5.05 & -4.83 &
-5.08 & $80_1\rightarrow 60_1$ & 5.90$_{-0.30}^{+0.30}$ &
5.80 & 5.60  & 5.80 \\
$100_1\rightarrow 100_1$ & -6.32$_{-0.32}^{+0.74}$ & -5.60 & -5.22
& -5.63 & $100_1\rightarrow 80_1$ & 6.71$_{-0.34}^{+0.34}$  &
6.54 & 6.29  & 6.54 \\
$120_1\rightarrow 120_1$ & -6.15$_{-0.74}^{+0.64}$  & -6.08 &
-5.53 & -6.15 & $120_1\rightarrow 100_1$ & 7.01$_{-0.35}^{+0.35}$
& 7.19
& 6.79  & 7.20 \\
$140_1\rightarrow 140_1$ & -- & -- & -- & 6.62 & $140_1\rightarrow
120_1$ & 8.12$_{-0.43}^{+0.63}$  & 7.80 & 7.18  & 7.81
\\ \hline
$22_1\rightarrow 00_1$ & 0.208 $_{-0.040}^{+0.010}$ & 0.21 &
0.20$^{a)}$ & 0.203 & $00_2\rightarrow 20_1$ &
0.166$_{-0.018}^{+0.018}$ & 0.16 & 0.27 & 0.01  \\
$\hspace*{0.6cm}\rightarrow  20_1$ & 0.250 $_{-0.018}^{+0.016}$ &
0.25 & 0.31 & 0.255 & $20_2\rightarrow 00_1$ &
0.090$_{-0.040}^{+0.010}$ & 0.16 & 0.26  & 0.082  \\
$\hspace*{0.6cm}\rightarrow  40_1$ & 0.063 $_{-0.004}^{+0.009}$ &
0.062 & 0.10 & 0.066 & $\hspace*{0.6cm}\rightarrow 20_1$ & -0.162
$_{-0.008}^{+0.071}$ &
 0.19 & -0.31  & 0.108 \\
$42_1\rightarrow 20_1$ & 0.22 $_{-0.05}^{+0.07}$ & 0.20 & 0.13 &
0.11 &  $\hspace*{0.6cm}\rightarrow  40_1$ & 0.27
$_{-0.08}^{+0.02}$ &  0.26 & 0.45  &
0.19  \\
$\hspace*{0.7cm}\rightarrow  40_1$ & 0.46 $_{-0.13}^{+0.08}$ &
0.38 & 0.45 & 0.27 & $40_2\rightarrow 40_1$ & -- &  -- &
--  & 0.13   \\
$32_1\rightarrow 20_1$ & 0.32(11) &
 -- & -- & 0.328  &  & & & &  \\
$\hspace*{0.7cm}\rightarrow  40_1$ & 0.235(6) &
 -- & -- & 0.226 &  & & & &  \\
\hline\hline
\end{tabular*}
\end{footnotesize}
\footnotesize{$^a)$ This matrix element was used to normalize the
results of the model calculations. }
\end{table}
\end{center}
\end{widetext}

To evaluate the degree of nonadiabaticity, manifested in the
reduced probabilities of $E2$-- transitions, in Table 5
theoretical ratios $R_{IK}$ has been compared with their
adiabatical values $R_{IK}^{adiab.}$ as well as experimental data
\cite{Baglin}-\cite{Browne}, \cite{Begzhanov, Reich, Gasten73}
which is determined as follows
\begin{eqnarray}\label{Eq.12}
R_{IK}=\frac{I_{\gamma}\left(IK\rightarrow
I_10_1\right)}{I_{\gamma}\left(IK\rightarrow
I_20_1\right)}\left(\frac{E_{\gamma}\left(IK\rightarrow
I_20_1\right)}{E_{\gamma}\left(IK\rightarrow
I_10_1\right)}\right)^5
\end{eqnarray}

\noindent where $I_{\gamma}\left(IK\rightarrow I_10_1\right)$-- is
intensity and $E_{\gamma}\left(IK\rightarrow I_10_1\right)$-- is
energy of $\gamma$-- transition.

\begin{widetext}
\begin{center}
\begin{table} [h]
{\textbf{\caption {The ratios of reduced probabilities of $E2$--
transitions $R_{_{IK}}=B(E2;I_iK_i\rightarrow
I_f0_1)/B(E2;I_iK_i\rightarrow I_{f}^{'}0_{1})$ in isotopes $Yb$
 }\label{TypoMSP}}}
\begin{tabular*} {0.90\textwidth}%
{@{\extracolsep{\fill}}ccccrrrr} \hline\hline
 A &  \ $I_iK_i$ \ & \ $I_fK_f$ \ & \ $I_{f}^{'}K_{f}^{'}$ \ &
    \multicolumn{2}{c}{Experiments} & \ \ Theory \ \ & \ \ Alaga \ \ \\ \hline
$^{170}Yb$ & $22_1^+$ & $20_1^+$ & $00_1^+$ & 1.77(8)\cite{Baglin} & & 1.86 & 1.43 \\
&  & $40_1^+$ & $20_1^+$ & 0.098(11)\cite{Baglin}  & & 0.043 & 0.050 \\
&  $32_1^+$ & $40_1^+$ & $20_1^+$ & 0.78(4)\cite{Baglin}  & & 0.75 & 0.40 \\
&  $52_1^+$ & $60_1^+$ & $40_1^+$ & 1.39(46)\cite{Baglin}  & & 1.50 & 0.57 \\
&  $72_1^+$ & $80_1^+$ & $60_1^+$ & 1.27(24)\cite{Baglin}  & & 2.42 & 0.67 \\
\hline
&  $20_2^+$ & $20_1^+$ & $00_1^+$ & 1.94(52)\cite{Baglin}  & & 1.1 & 1.43 \\
&  $40_2^+$ & $40_1^+$ & $20_1^+$ & --  &  & 1.82 & 0.91 \\
&  $60_2^+$ & $60_1^+$ & $40_1^+$ & 3.73(90)\cite{Baglin}  & & 1.67 & 0.81 \\
&  $80_2^+$ & $80_1^+$ & $60_1^+$ & 10.7(19)\cite{Baglin}  & & 2.45 & 0.77 \\
&  $100_2^+$ & $100_1^+$ & $80_1^+$ & 25.3(85)\cite{Baglin}  & & 29.2 & 0.74 \\
&  $120_2^+$ & $120_1^+$ & $100_1^+$ & 29.9(71)\cite{Baglin}  & & 2.0 & 0.73 \\
\hline
&  $20_3^+$ & $20_1^+$ & $00_1^+$ & 1.81(11)\cite{Baglin}  &  & 2.5 & 1.43 \\
&  & $40_1^+$ & $20_1^+$ & 3.00(15)\cite{Baglin}  & & 2.3 & 1.80 \\
&   & $40_1^+$ & $00_1^+$ & 5.43(18)\cite{Baglin}  & & 5.75 & 2.57 \\
&  $20_4^+$ & $40_1^+$ & $00_1^+$ & 4.0(2)\cite{Baglin}  & & 3.9 & 2.57 \\
\hline \hline

$^{172}Yb$ & $22_1^+$ & $20_1^+$ & $00_1^+$ & 1.62(12) \cite{Reich} &  1.71(84) \cite{Singh} &   1.59 & 1.43   \\
 &   & $40_1^+$ & $20_1^+$ & 0.056(5) \cite{Reich} & 0.056(20) \cite{Singh} &  0.066 & 0.072   \\
 &  $32_1^+$ & $40_1^+$ & $20_1^+$ & 0.52(4) \cite{Reich} & 0.56(3) \cite{Singh} &  0.48 & 0.40   \\
 &  $42_1^+$ & $40_1^+$ & $20_1^+$ &  & 3.35(69) \cite{Singh} & 6.0 & 2.94
\\ \hline
 &  $22_2^+$ & $20_1^+$ & $00_1^+$ & 1.69(21) \cite{Reich} & 1.55(30) \cite{Singh} &  1.61 & 1.43  \\
 &   & $40_1^+$ & $20_1^+$ & 0.015(8) \cite{Reich} & 0.064(38) \cite{Singh} &  0.058 & 0.072   \\
 &  $32_2^+$ & $40_1^+$ & $20_1^+$ & 0.163(56) \cite{Reich} & 0.409(77) \cite{Singh} &  0.504 & 0.40    \\
 &  $42_2^+$ & $40_1^+$ & $20_1^+$ &  & 4.11(17) \cite{Singh} & 3.89 & 2.94 \\
 &  $52_2^+$ & $60_1^+$ & $40_1^+$ &  & $<$ 1.10 \cite{Singh} & 0.82 & 0.57 \\
  \hline
 &  $20_2^+$ & $20_1^+$ & $00_1^+$ &
 3.71(24) \cite{Reich} &  2.88(36) \cite{Singh} &  1.80 & 1.43   \\
 &   & $40_1^+$ & $20_1^+$ & 2.70(38) \cite{Reich} & 2.61(11) \cite{Singh} &  3.20 & 1.80 \\
 &  $40_2^+$ & $40_1^+$ & $20_1^+$ & & 6.78(1.36) \cite{Singh} & 1.56 & 0.91
\\  \hline \hline

$^{174}Yb$ & $22_1^+$ & $20_1^+$ & $00_1^+$ & $>$ 0.49\cite{Begzhanov, Gasten73} &  2.4(5)\cite{Browne} & 1.59 & 1.43  \\
 &   & $40_1^+$ & $20_1^+$ & 0.167(75)\cite{Begzhanov, Gasten73} & 0.256(92)\cite{Browne} & 0.055 & 0.072   \\
 &  $32_1^+$ & $40_1^+$ & $20_1^+$ & $>$ 0.325\cite{Begzhanov, Gasten73} & 0.67(13)\cite{Browne} & 0.49 & 0.40  \\
 &  $42_1^+$ & $40_1^+$ & $20_1^+$ & 4.83(5)\cite{Begzhanov, Gasten73} & 4.77(63)\cite{Browne} & 3.75 & 2.94   \\
 &   & $60_1^+$ & $40_1^+$ & $\leq$ 0.14\cite{Begzhanov, Gasten73}  &  &  0.09 & 0.086 \\
 \hline
 &  $20_2^+$ & $20_1^+$ & $00_1^+$ & $>$ 8.8\cite{Begzhanov, Gasten73} & $>$ 9.7\cite{Browne} & 2.63 & 1.43  \\
 &   & $40_1^+$ & $20_1^+$ & 3.45(5)\cite{Begzhanov, Gasten73} & 2.9(2)\cite{Browne} & 6.1 & 1.80   \\
 &  $40_2^+$ & $40_1^+$ & $20_1^+$ & 7.82(26)\cite{Begzhanov, Gasten73} & 11.8(2.5)\cite{Browne} &  6.21 & 0.91   \\
 &   & $60_1^+$ & $40_1^+$ & 1.50(22)\cite{Begzhanov, Gasten73} & 0.59(7)\cite{Browne}  &  11.3 & 1.75  \\
\hline
 &  $20_4^+$ & $20_1^+$ & $00_1^+$ & $>$ 1.7\cite{Begzhanov, Gasten73} & 1.99(27)\cite{Browne} &  1.54 & 1.43   \\
 &   & $40_1^+$ & $20_1^+$ & 1.24(16)\cite{Begzhanov, Gasten73} & 1.12(12)\cite{Browne}  &   2.11 & 1.80   \\
 \hline\hline
\end{tabular*}
\end{table}
\end{center}
\end{widetext}

In $^{170}Yb$, experimental ratio $R_{I0_2}$ for $E2$--
transitions from states $K^{\pi}=0_2^+$ band differ from adiabatic
theory considerably (10-40 times). This is associated with mixing
$0_2^+$ and $2_1^+$ bands. An important question can raised in
this regard. Why do the ratios $R_{I2_1}$ for transitions from
$2_1^+$ bands differ not so strongly from adiabatic theory with
respect to $R_{I0_2}$?

This results can be explained by the fact that the matrix element
$m_{2_1}$ is greater about 10 times than that of $m_{0_2}$ (see
Table 2). One may see from this comparison that the mixing effect
of states low-lying bands plays a crucial role which considerably
demonstrates that $E2$-- transitions even in low values of angular
momentum $I$.

\section{Conclusion}

In the present work, non-adiabatic effects in energies and
electric characteristics of excited states are studied within the
phenomenological model which taking into account Coriolis mixing
of all experimentally known rotational bands with $K^{\pi}< 3^+$.

The energy and structure of wave functions of excited states are
calculated. And also the reduced probabilities of $E2$-- transitions
 is calculated. The ratio of $E2$-- transitions
probability from $K^{\pi}=0_m^+$ and $2_{\ell}^+$ bands are
calculated and compared with experimental data which gives the
satisfactory result.

{\bf\it  If matrix elements of $E2$-- transitions $m_K$ one of two
strongly mixing bands $K$ is less than matrix element of
$m_{_{K'}}$ of $K'$ $(m_{_K}<m_{_{K'}})$, then the difference in
the ratio $R_{_{IK}}$ for the first band $K$ from Alaga rule is
bigger than the difference in $R_{IK'}$ from Alaga rule.} In other
words, if $m_{_K}<m_{_{K'}}$, nonadiabaticity in the ratio
$R_{_{IK}}$ is stronger than that of $R_{_{IK'}}$.

\begin{acknowledgments}
This work has been financial supported by the MOHE, Fundamental
Research Grant Scheme FRGS13-074-0315. We thank the Islamic
Development Bank (IDB) for supporting this work under scholarships
 37/IRQ/P30.

\end{acknowledgments}


\begin{figure}[h!] \label{Fig.2}
  \includegraphics[width=0.70\textwidth]{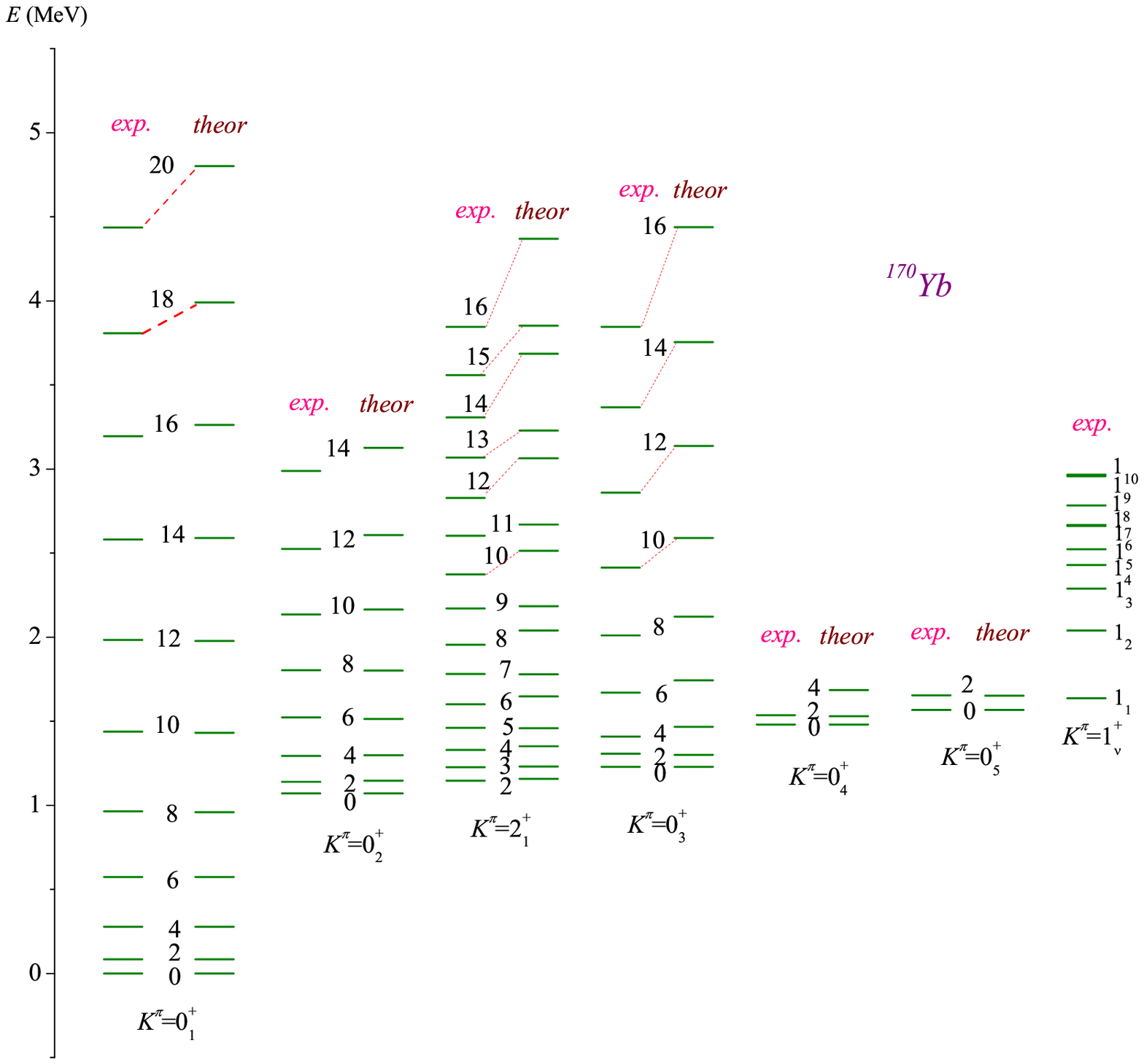}\\
  \caption{(Color online) Comparison of the calculated and
experimental energy spectra of positive-parity states for
$^{170}Yb$. }
\end{figure}

\begin{figure}[h!] \label{Fig.3}
  \includegraphics[width=0.60\textwidth]{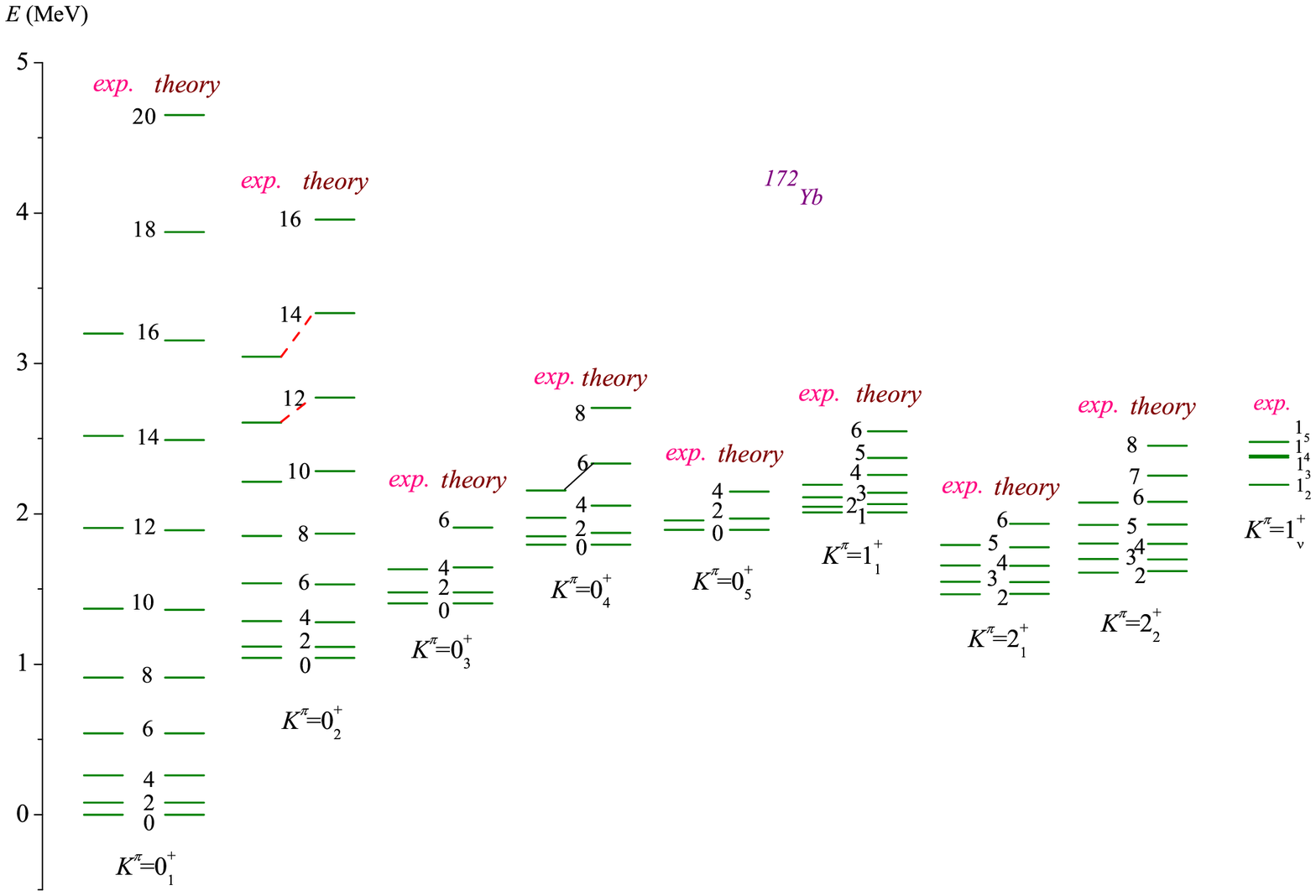}\\
  \caption{(Color online) Comparison of the calculated and
experimental energy spectra of positive-parity states for
$^{172}Yb$. }
\end{figure}

\begin{figure}[h!] \label{Fig.4}
  \includegraphics[width=0.65\textwidth]{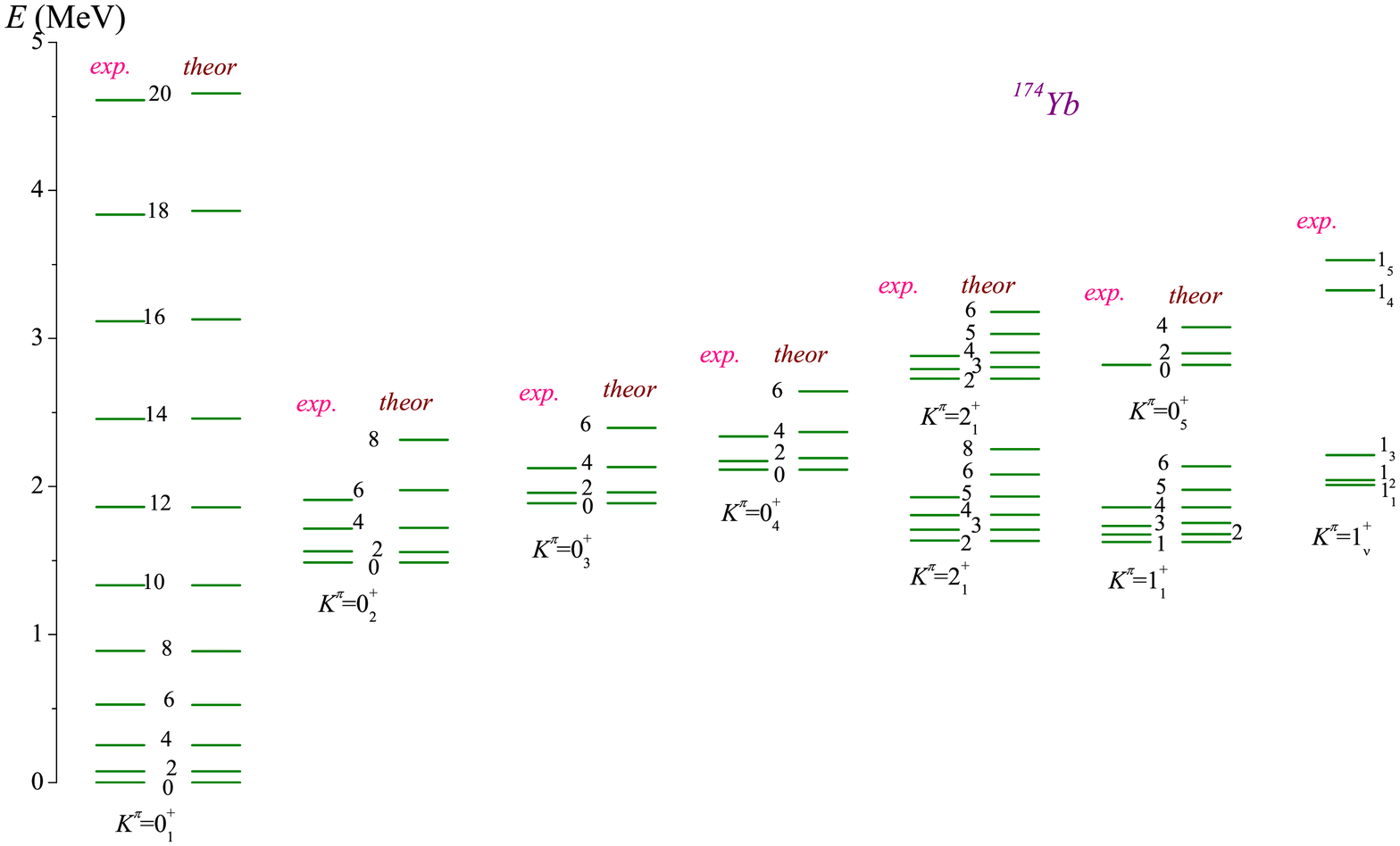}\\
  \caption{(Color online) Comparison of the calculated and
experimental energy spectra of positive-parity states for
$^{174}Yb$. }
\end{figure}

\begin{figure}[h!] \label{Fig.5}
  \includegraphics[width=0.50\textwidth]{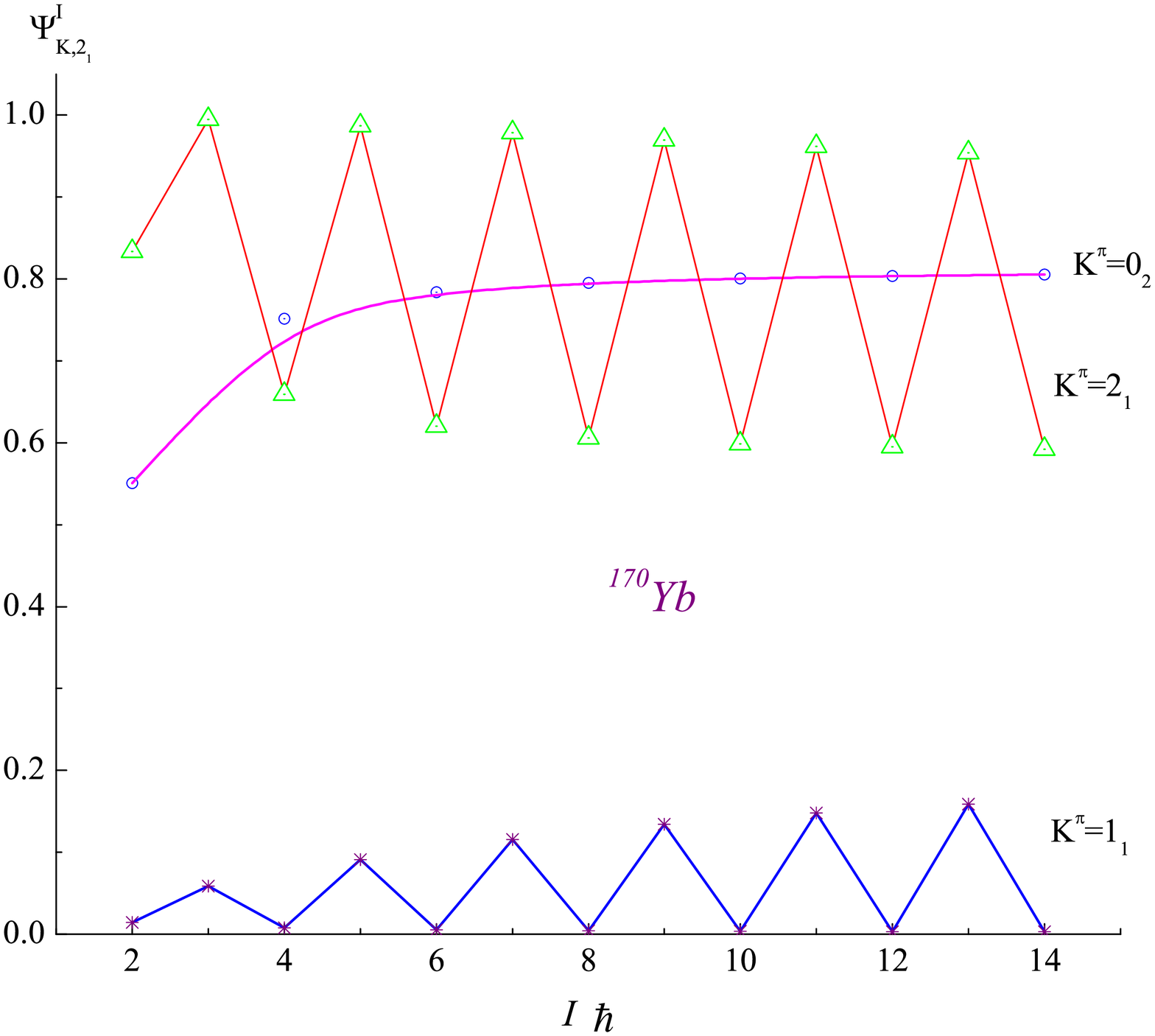}\\
  \caption{(Color online) Structure of the wave-functions of
$2_1^+$-- band states for $^{170}Yb$. }
\end{figure}

\begin{figure}[h!] \label{Fig.6}
  \includegraphics[width=0.50\textwidth]{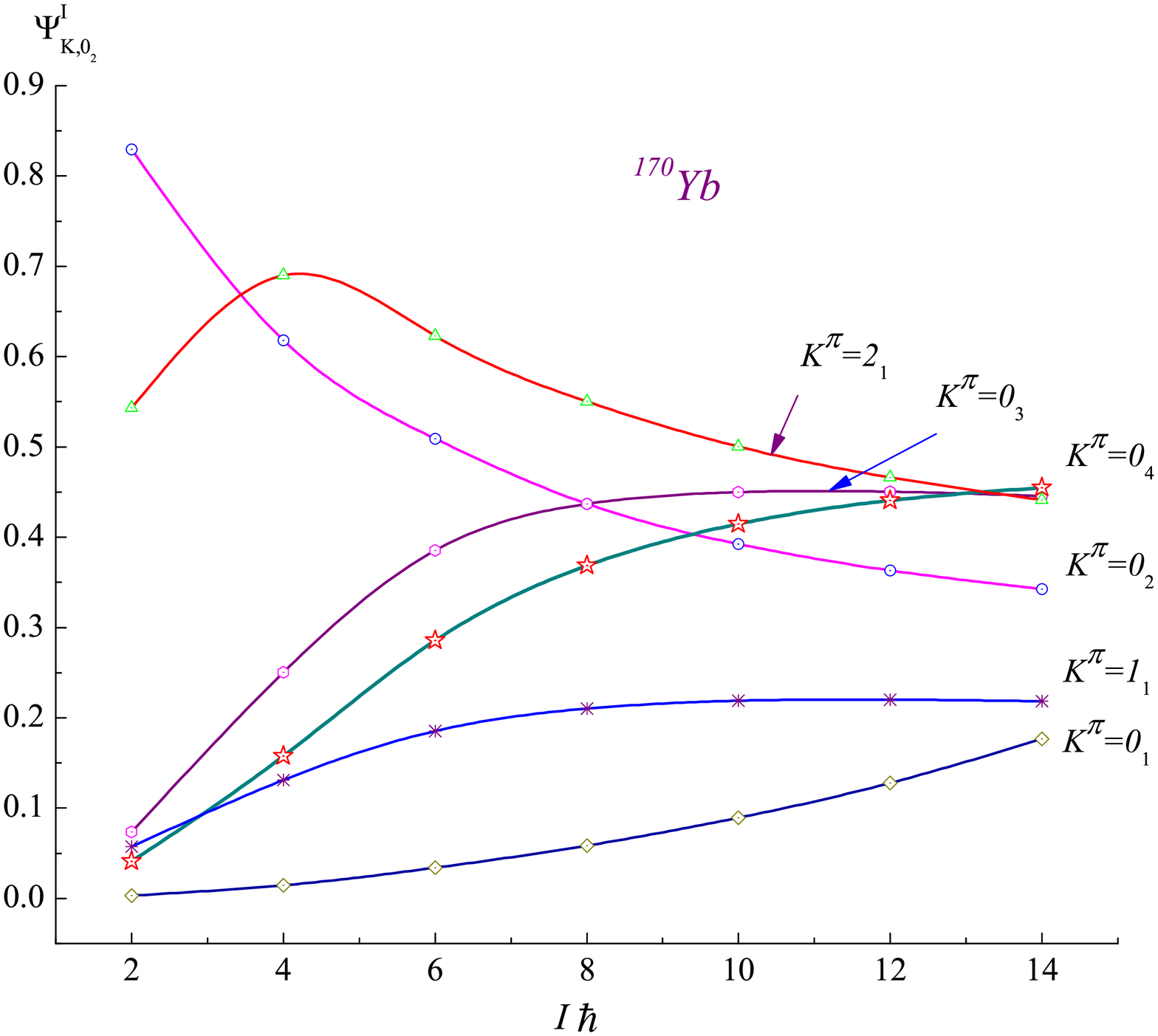}\\
  \caption{(Color online) Structure of the wave-functions of
$0_2^+$-- band states for $^{170}Yb$. }
\end{figure}

\begin{figure}[h!] \label{Fig.7}
  \includegraphics[width=0.5\textwidth]{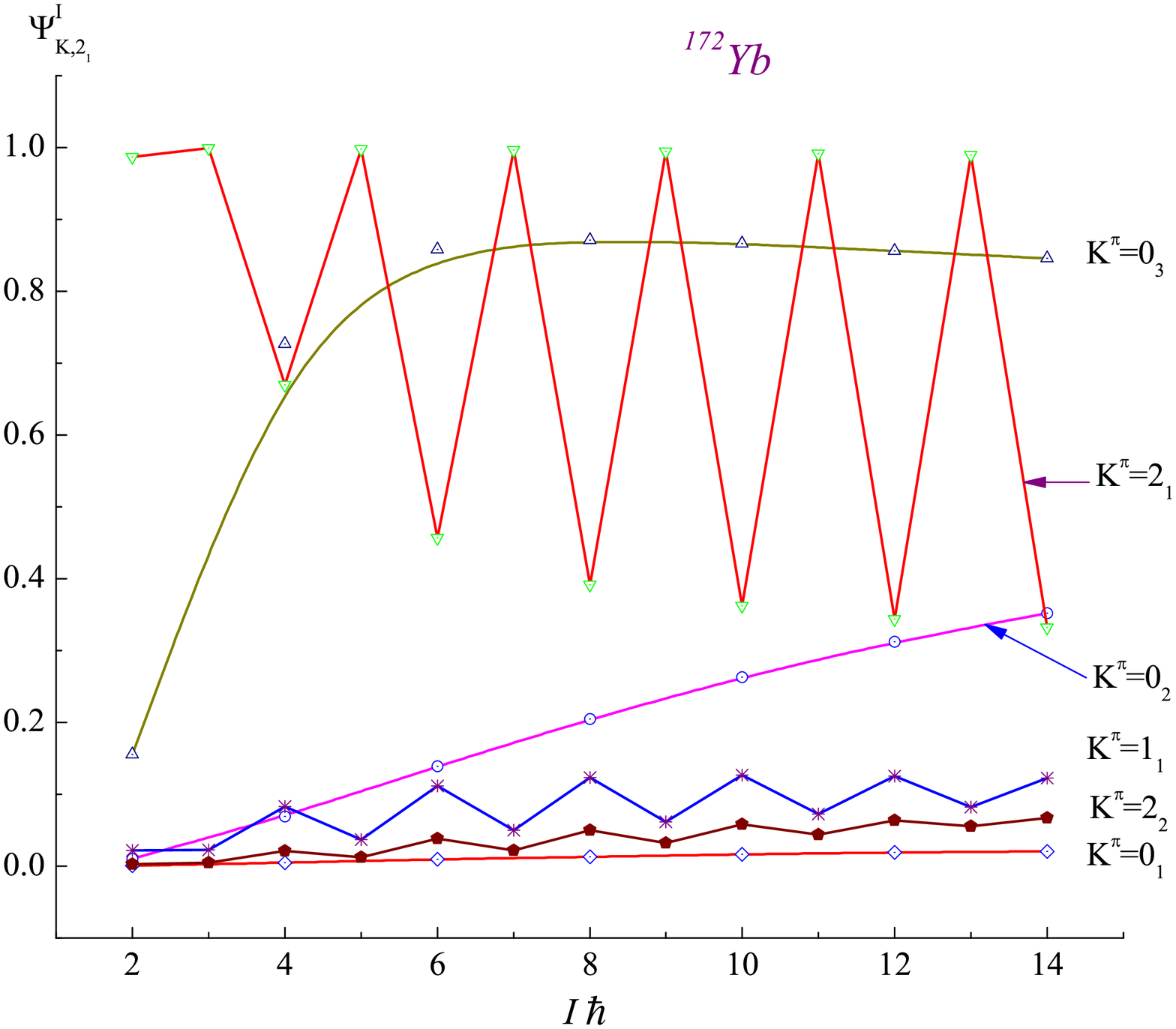}\\
  \caption{(Color online) Structure of the wave-functions of
$2_1^+$-- band states for $^{172}Yb$. }
\end{figure}

\begin{figure}[h!] \label{Fig.8}
  \includegraphics[width=0.5\textwidth]{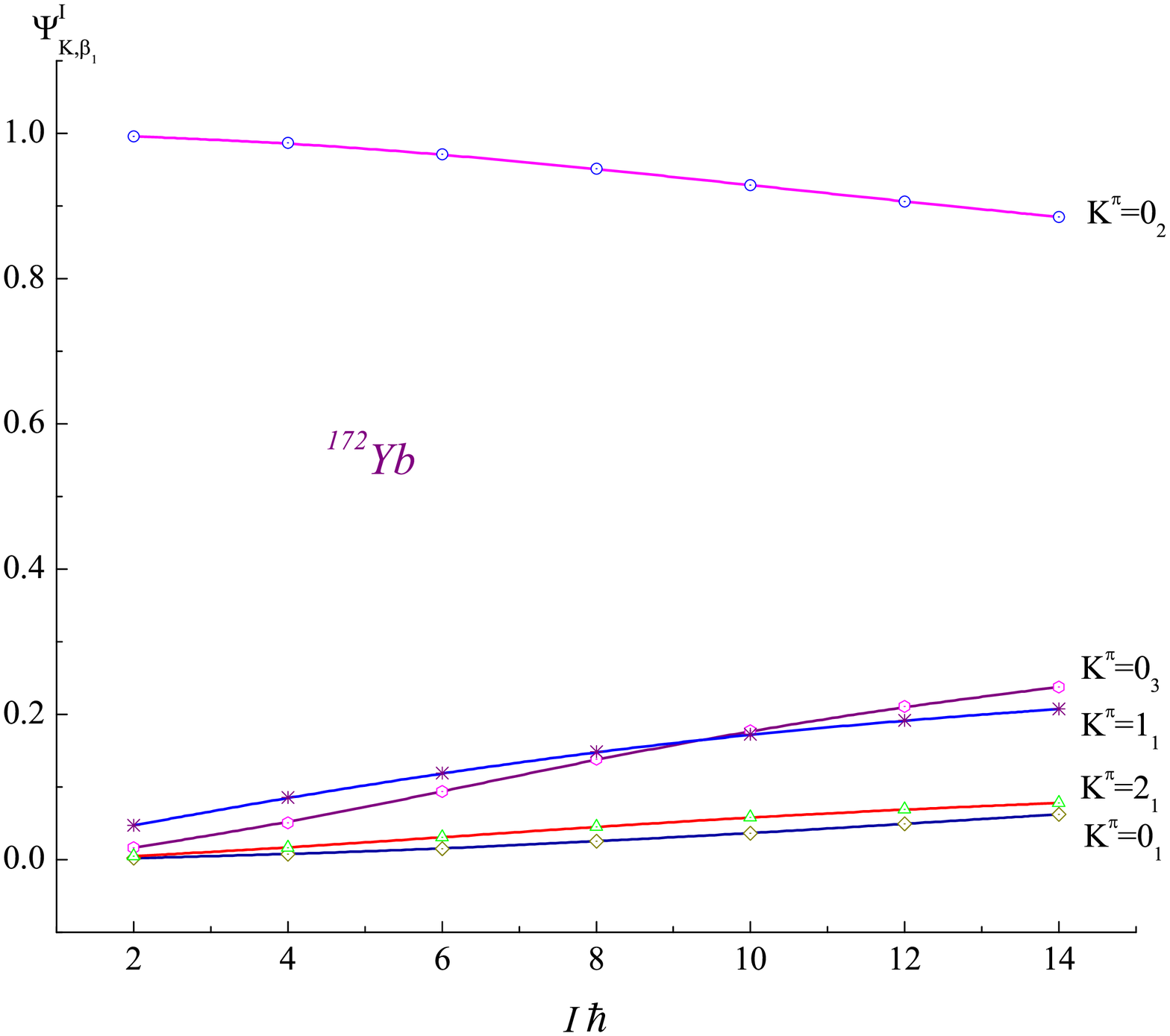}\\
  \caption{(Color online) Structure of the wave-functions of
$0_2^+$-- band states for $^{172}Yb$. }
\end{figure}

\begin{center}
\begin{figure}[htb] \label{Fig.8}
  \includegraphics[width=0.5\textwidth]{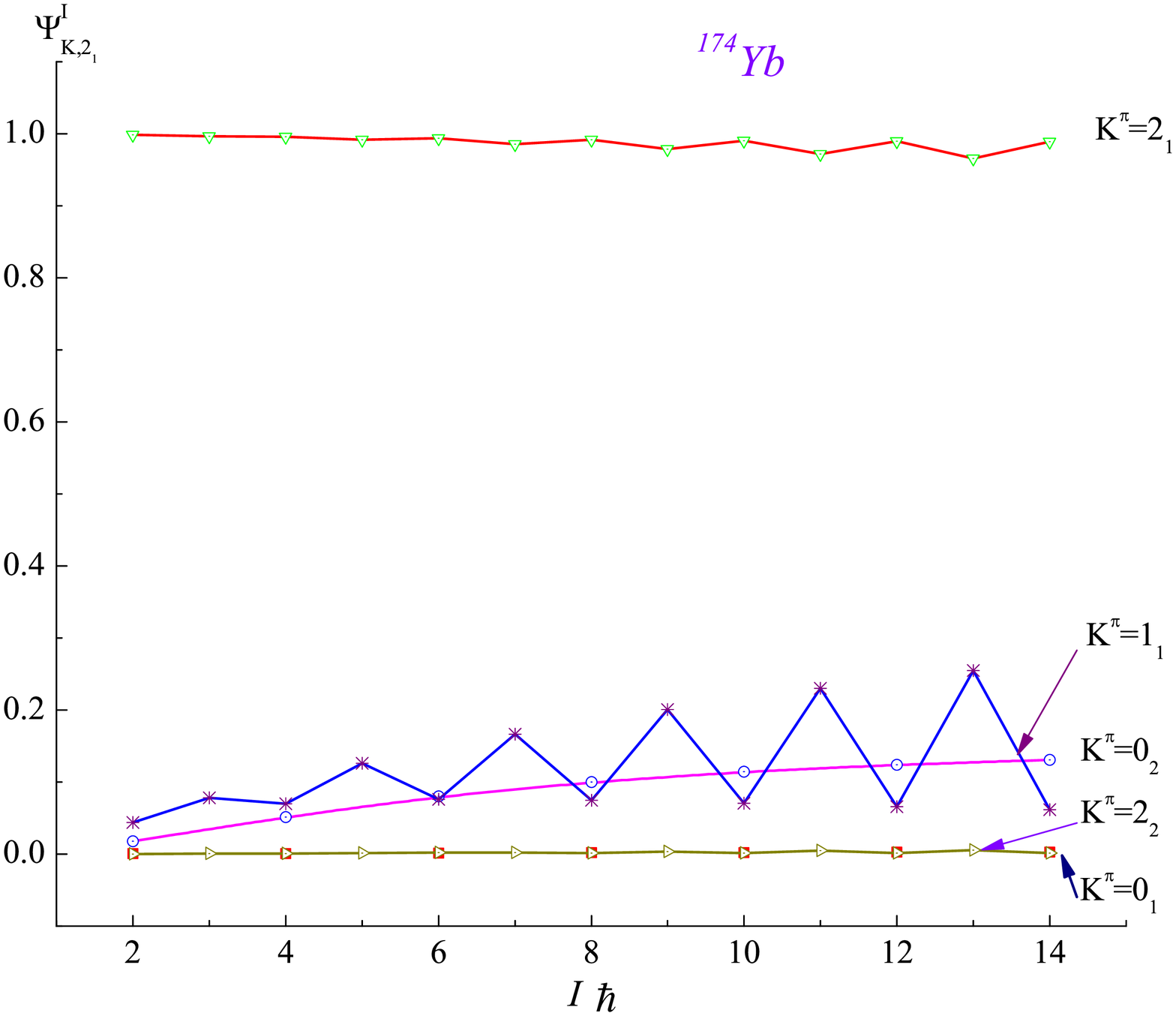}\\
  \caption{(Color online) Structure of the wave-functions of
$2_1^+$-- band states for $^{174}Yb$. }
\end{figure}
\end{center}

\begin{center}
\begin{figure}[htb] \label{Fig.9}
  \includegraphics[width=0.5\textwidth]{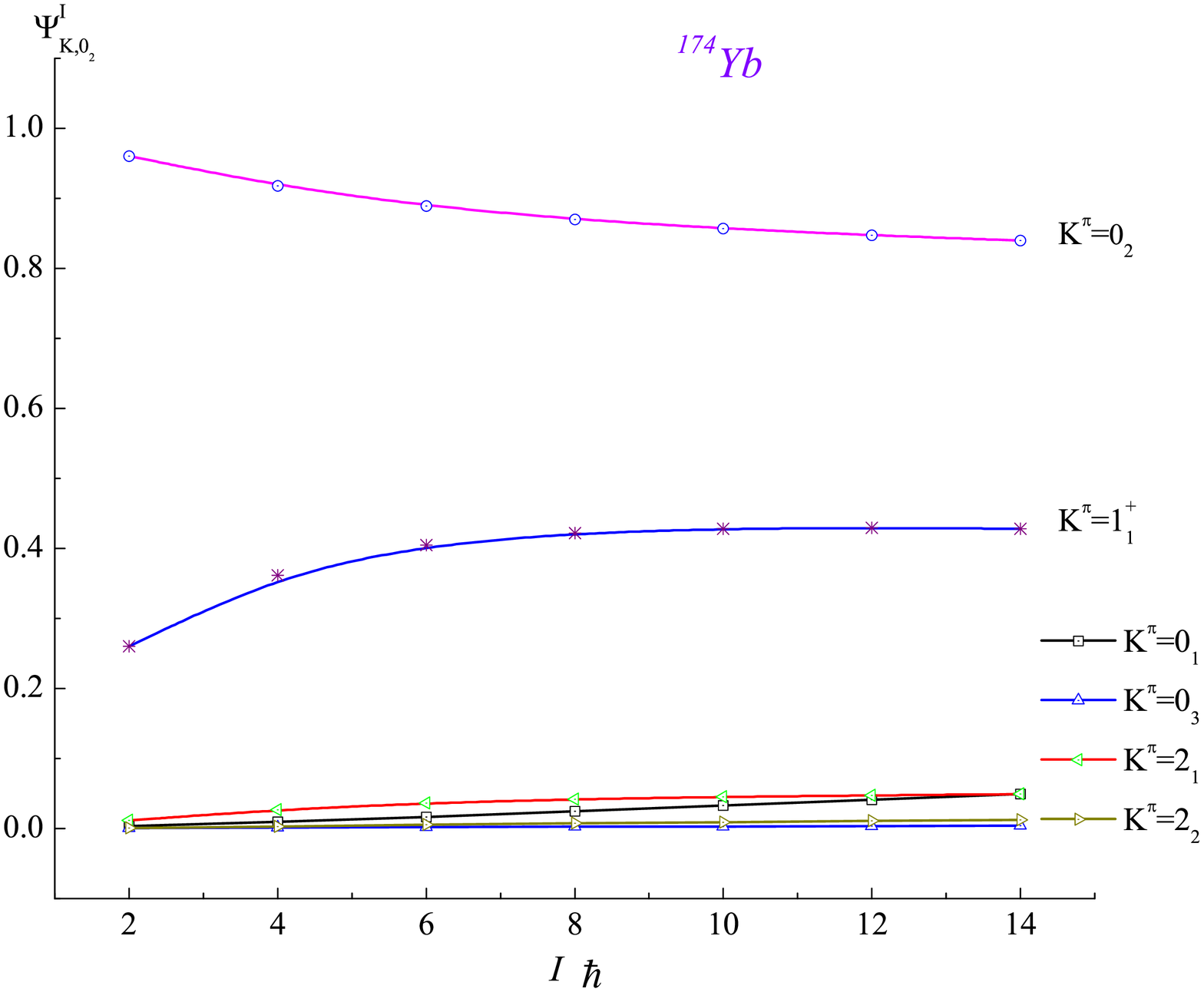}\\
  \caption{(Color online) Structure of the wave-functions of
$0_2^+$-- band states for $^{174}Yb$. }
\end{figure}
\end{center}

\end{document}